\documentclass{article}
\usepackage{graphicx}
\usepackage{hyperref}

\title{Cross correlations of the American baby names}

\author{Paolo Barucca$^1$ \qquad Jacopo Rocchi$^1$ \qquad Enzo Marinari$^{1,2}$\\ \qquad Giorgio Parisi$^{1,2}$ \qquad Federico Ricci-Tersenghi$^{1,2}$}

\begin{document}
\maketitle
\begin{abstract}
{The quantitative description of cultural evolution is a challenging task. The most difficult part of the problem is probably to find the appropriate measurable quantities that can make more quantitative such evasive concepts as, for example, dynamics of cultural movements, behavior patterns and traditions of the people. A strategy to tackle this issue is to observe particular features of human activities, i.e. cultural traits, such as names given to newborns. We study the names of babies born in the United States of America from 1910 to 2012. Our analysis shows that groups of different correlated states naturally emerge in different epochs, and we are able to follow and decrypt their evolution. While these groups of states are stable across many decades, a sudden reorganization occurs in the last part of the twentieth century. We think that this kind of quantitative analysis can be possibly extended to other cultural traits: although databases covering more than one century (as the one we used) are rare, the cultural evolution on shorter time scales can be studied thanks to the fact that many human activities are usually recorded in the present digital era.}
\end{abstract}

\footnotetext[1] { Dipartimento di Fisica, Sapienza Universit{\'a} di Roma, P.le A. Moro 2, I-00185 Roma, Italy}
\footnotetext[2] {INFN, Sezione di Roma 1 and IPFC-CNR, UOS di Roma, P.le A. Moro 2, I-00185 Roma, Italy}

Cultural traits can be considered as the fundamental blocks of the culture of a community. They are characterized by a much shorter time scale than cultural movements: while changes in cultural movements may occur over decades or centuries, changes in cultural traits may be observed from a daily to an yearly basis, depending on the trait. This feature allows one to start from measured data to analyse their reciprocal influence: a cultural trait may promote or prevent the popularity rise of others, a past cultural trait may have an influence on current and future ones and finally the rise or fall of a cultural trait in a certain area may influence cultural traits in other areas.

Many cultural traits have been studied in the past. Among them are skirt lengths \cite{belleau1987cyclical}, pop songs \cite{bentley2007regular}, dog breeds \cite{herzog2004random} and pottery decorations in the archaeological record \cite{neiman1995stylistic}. Also keywords in academics vocabulary have been the focus of recent interest \cite{bentley2008random}. Data about names given to newborns have been investigated for similar reasons \cite{krawczyk2014simmel, xi2014cultural}. Names come and go in the society, as any other cultural trait. Most of them have a popularity peak and then disappear. They carry important information on the transformation of the social structure. Several quantitative approaches to these problems have been proposed, and we will briefly describe them in the following. When compared to other relevant traits, names appear very appropriate to study cultural changes, since the success of a name mainly depends on the influence that the surrounding culture wields on the parents of the newborns. Other traits suffer, for example, the influence of external forces, such as that due to the producers which may artificially shape the tastes of consumers. This is particularly evident in the fashion market and in the music market \cite{berger2012karen}.

The frequency distribution of names given to newborns have fat tails, typical of many physical problems \cite{jensen1998self}. Fat tailed distributions can be generated by different mechanisms \cite{albert2002statistical, feldman1981further, neiman1995stylistic} and also in the case of names they have been given several explanations. A scale free network was used to study a fashion diffusion process where each node could take one out of many values \cite{krawczyk2014simmel}, imitating "popular" nodes and avoiding "non-popular" nodes. Moreover, a stochastic model for cultural evolution was proposed, where names were chosen according to both individual preferences and social influence \cite{xi2014cultural}. These different mechanisms are all able to reproduce a fat-tailed distribution and were shown to recover several features of the real data. The popularity of a name was also shown to be correlated with the popularity of similar names in previous years \cite{berger2012karen}. Furthermore, names were analized in terms of activation and inhibition processes \cite{zanette2012dynamics}, in order to explain their popularity rise and fall and the duration of the rise time and fall time of the popularity of names were found to be correlated \cite{acerbi2012logic} and \cite{berger2009adoption}.

The mechanisms behind the spreading of cultural traits is still debated. The original hypothesis of Simmel \cite{Simmel} was that a fashion arises because individuals of lower social status copy those of perceived higher status. This is the idea used for the analysis done in \cite{krawczyk2014simmel}. This approach is different from the \emph{neutral model} proposed in \cite{hahn2003drift}, where naming was considered in close connection with the infinite-allele model of population genetics with a random genetic drift. A preference model of fashion \cite{acerbi2012logic} where individuals can copy each other�s preferences in addition to traits themselves, was said to better reproduce the empirical features of the American baby names. These previous studies on names were mainly focused on global distributions, but not at the relations between local distributions of names in different states (i.e. distributions in single states).
We believe that much can be learned, for example, from the relations between local distributions of names in different states. Our main working hypothesis is that local changes of names convey a large body of information on the mutual cultural influences that communities (states) wield on each other. 

\begin{figure*}
\centerline{\includegraphics[width=1\textwidth]{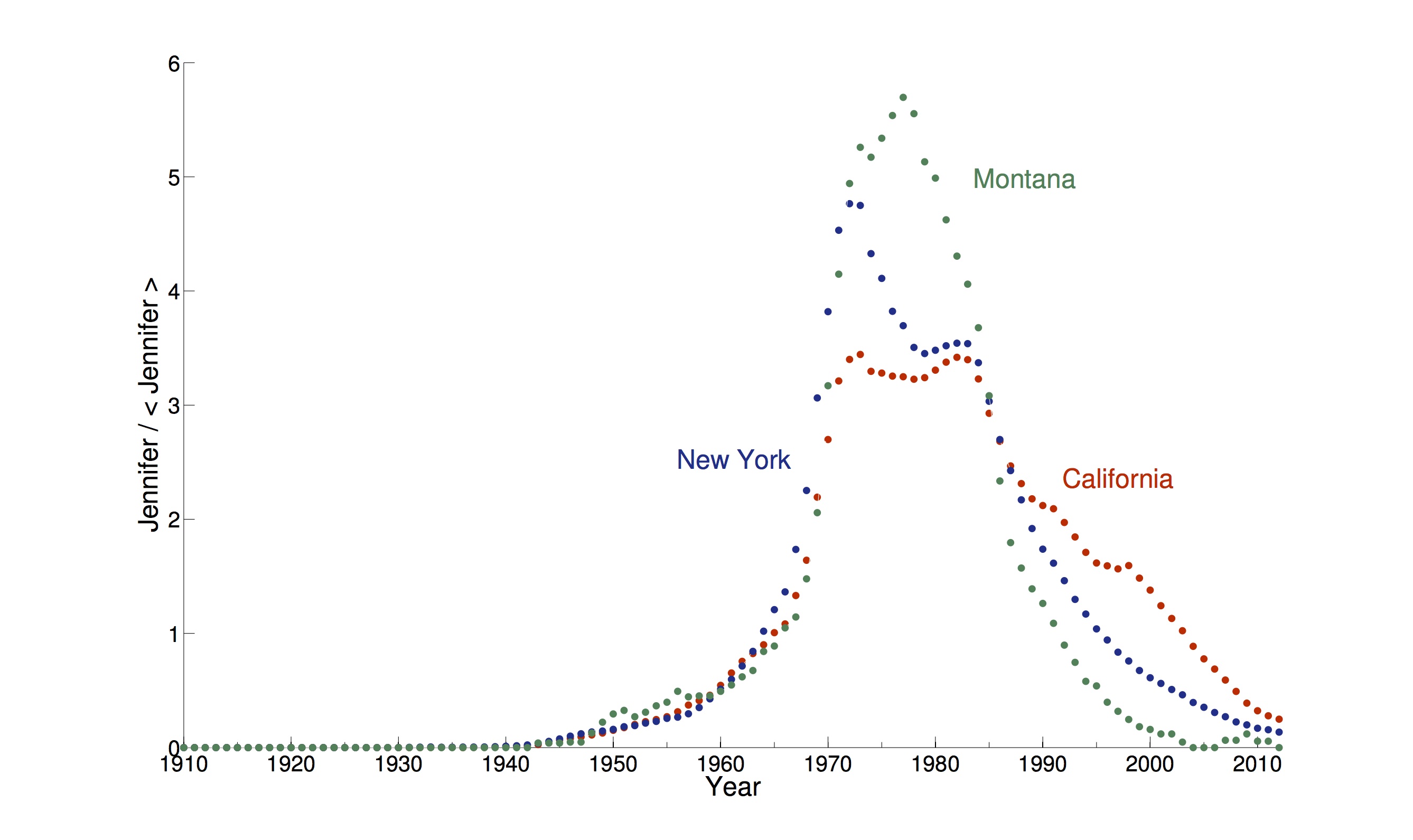}}
\caption{Number of newborns called Jennifer in the states of California, Montana and New York, divided by its average values in these three states, as a function of time. 
}\label{fig1}
\end{figure*}

\begin{figure*}
\centerline{\includegraphics[width=1\textwidth]{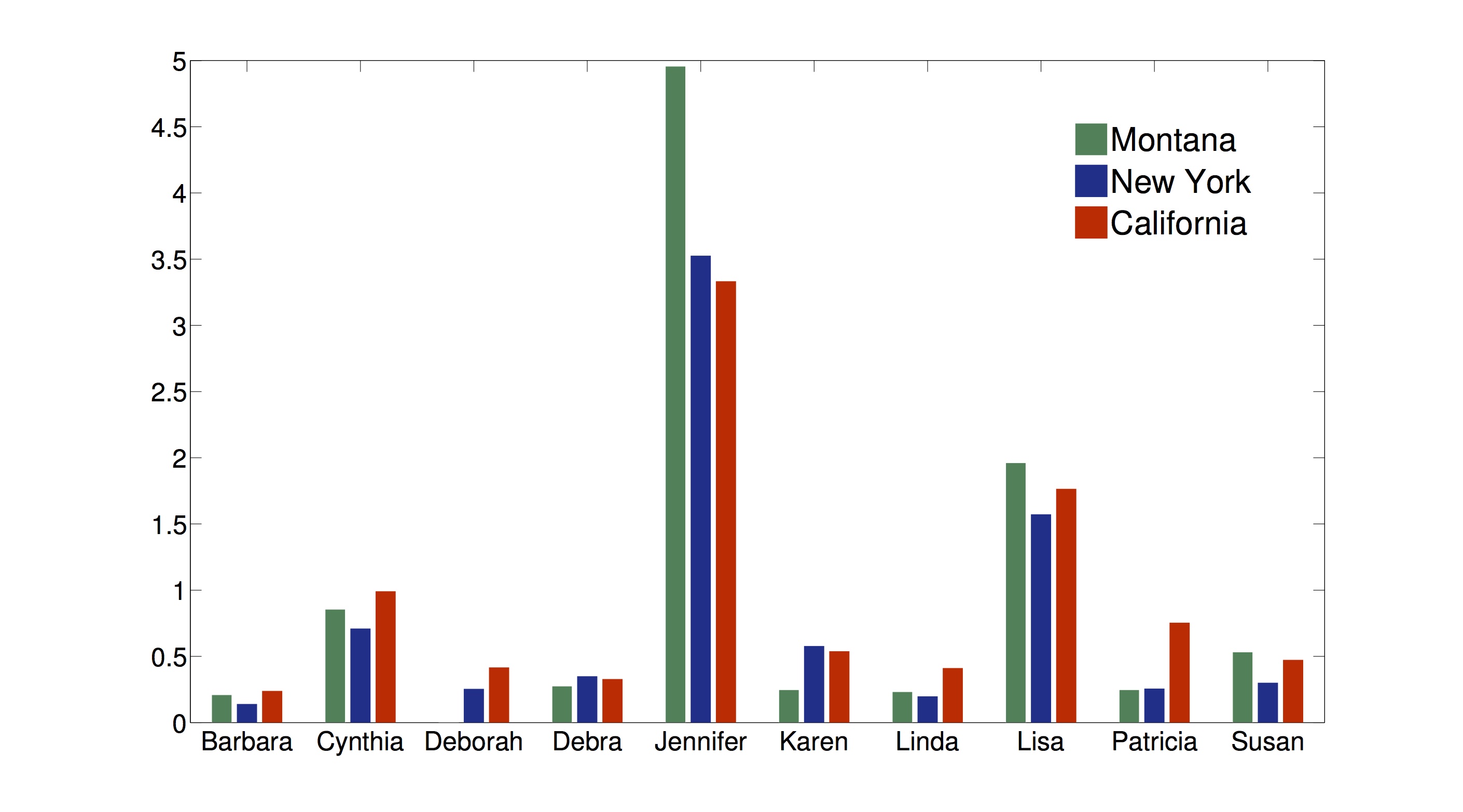}}
\caption{Histogram of the occurrences of the ten most popular names given in 1980 in the states of California, Montana and New York. Normalization is as in Fig. 1}\label{fig2}
\end{figure*}

We focus our correlation analysis on different states of the United States of America during the XX century. Statistics on names given to newborns in the United States can be downloaded from the web-page of the US Social Security Administration \cite{ssa}. Different states have different popularity spreading curves for each name, Fig. \ref{fig1} (many of the common names rise and fade with a very similar behaviour). The overlap between these curves could be used to describe the similarities between states. Instead of considering these overlaps in time, we consider the correlations between states on a yearly basis, by studying the whole distribution of baby names in every state, Fig. \ref{fig2}. This analysis gives robust results, as we will show in the following.

\section{Methods}
For each available year (that ranges, in the US Social Security Administration (SSA) archives \cite{ssa}, from 1910 to 2012) we study how names given in a state $i$ are correlated to names given in a state $j$. The distribution of these names have already been analized \cite{li2012analyses} and it is further described in the Supplementary Information. For each pair of states $i$ and $j$, with $i,j=1,\ldots, M=51$ (the federal district of Washington D.C. is considered by itself) we compute a correlation coefficient $C_{ij}$ , computed as follow.  Let us consider a generic year $y$ and let $n_S(q)$ be the number of girls named $q$ born in the state $S$ in the year $y$ (we limit ourself to describe the girls case as we have verified that analysing baby boys names leads to the same conclusions). In each year, we have a $N_f \times M$ rectangular matrix, whose entries $n_S(q)$ are the occurrences of the baby girl names, with $q=1,\ldots, N_f$ and $S=1,\ldots,M$, $N_f=19492$ being the number of different girl names, and $M=51$ the number of states. If the name $q$ appears less than 5 times in the state $S$ in the year $y$, then $n_S(q)$, because the information provided by SSA does not include names that occur less than 5 times. These matrices are sparse, and only an average of $3\%$ of the entries are different from zero. The frequency $f_S(q)$ of the name $q$ in the state $S$ is given by
\begin{equation}
f_S(q)=\frac{n_S(q)}{\sum_{q=1}^{N_f} n_S(q)}\;.
\label{eq:uno}
\end{equation}
The average frequency of the name $q$ over the states is
\begin{equation}
\overline{f}(q)=\frac{1}{M}\sum_{S=1}^M f_S(q)\;,
\label{eq:due}
\end{equation}
where $M=51$ is the total number of states.
We can define the quantities 
\begin{equation}
\tilde{f}_S(q)=f_S(q)-\overline{f}(q)\;,
\label{eq:tre}
\end{equation}
which take into account fluctuations of the frequencies of the names over the states. The average of $\tilde{f}_S(q)$ over all the names is zero in each state $S$, as can be explicitly seen from their definition:
\begin{equation}
\left< \tilde{f}_S(q) \right>_q = \frac{1}{N_f}\sum_{q=1}^{N_f} \tilde{f}_S(q) = \frac{1}{N_f}-\frac{1}{N_f}=0\:,
\end{equation}
given that in each state $S$, $\sum_{q} f_S(q) =1$. In order to compute how the names in the state $i$ are correlated with the names in the state $j$, we computed, year after year, the Pearson correlation between the variables $\tilde{f}_i$ and $\tilde{f}_j$, which is the $M \times M$ square matrix
\begin{equation}
C_{ij}=\frac{\sum_{n=1}^{N_f}\tilde{f}_i(n) \tilde{f}_j(n) }{\sqrt{ \sum_{n=1}^{N_f} \tilde{f}^2_i(n)  } \sqrt{ \sum_{n=1}^{N_f} \tilde{f}^2_j(n)  } }\:.
\label{eq:cij}
\end{equation}
This matrix can be used to capture the emergence of complex correlations between clusters of states and to study their evolution in time. However, separating the real information from the noise is a non-trivial problem. Similar issues have already been faced in biological problems \cite{sporns2004organization}, \cite{luo2006application}, \cite{luo2007constructing}, \cite{jalan2010random}, in financial stock-markets \cite{laloux1999noise}, \cite{plerou1999universal}, \cite{plerou2002random}, \cite{conlon2009cross} as well as in Internet traffic analysis \cite{barthelemy2002large} and in the statistics of atmospheric correlations \cite{santhanam2001statistics}. The main point is that even though the empirical correlation matrix is noisy it does have stable properties. We checked these properties, such as the eigenvalue spectrum and eigenvector localization, and compared them with a null hypothesis, i.e. random matrix properties, to identify the stable, non solely random, data correlations in a given period.  

Here we apply two general method for the analysis of correlation matrices, i.e. Principal Component Analysis (PCA) and hierarchical clustering.  PCA consists in selecting the eigenvectors corresponding to the largest eigenvalues of the cross-correlation matrix . This relies on the hypothesis that smaller eigenvalues are related to noise while larger ones are more related to the system dynamics.  Hierarchical clustering, on the other hand, starts from $M$ clusters formed of one state each and allows to set up a hierarchy of clusters by merging clusters according to their distances, that can be defined in several way from the correlations.  Both these widely known methods shows the same results that are stable in sexes, i.e. hold both for male and female names, year after year and for different metrics in the hierarchical clustering algorithm. These methods are further described in the Supplementary Material. 

\section{Results}

\begin{figure*}
\centerline{\includegraphics[width=0.9\textwidth]{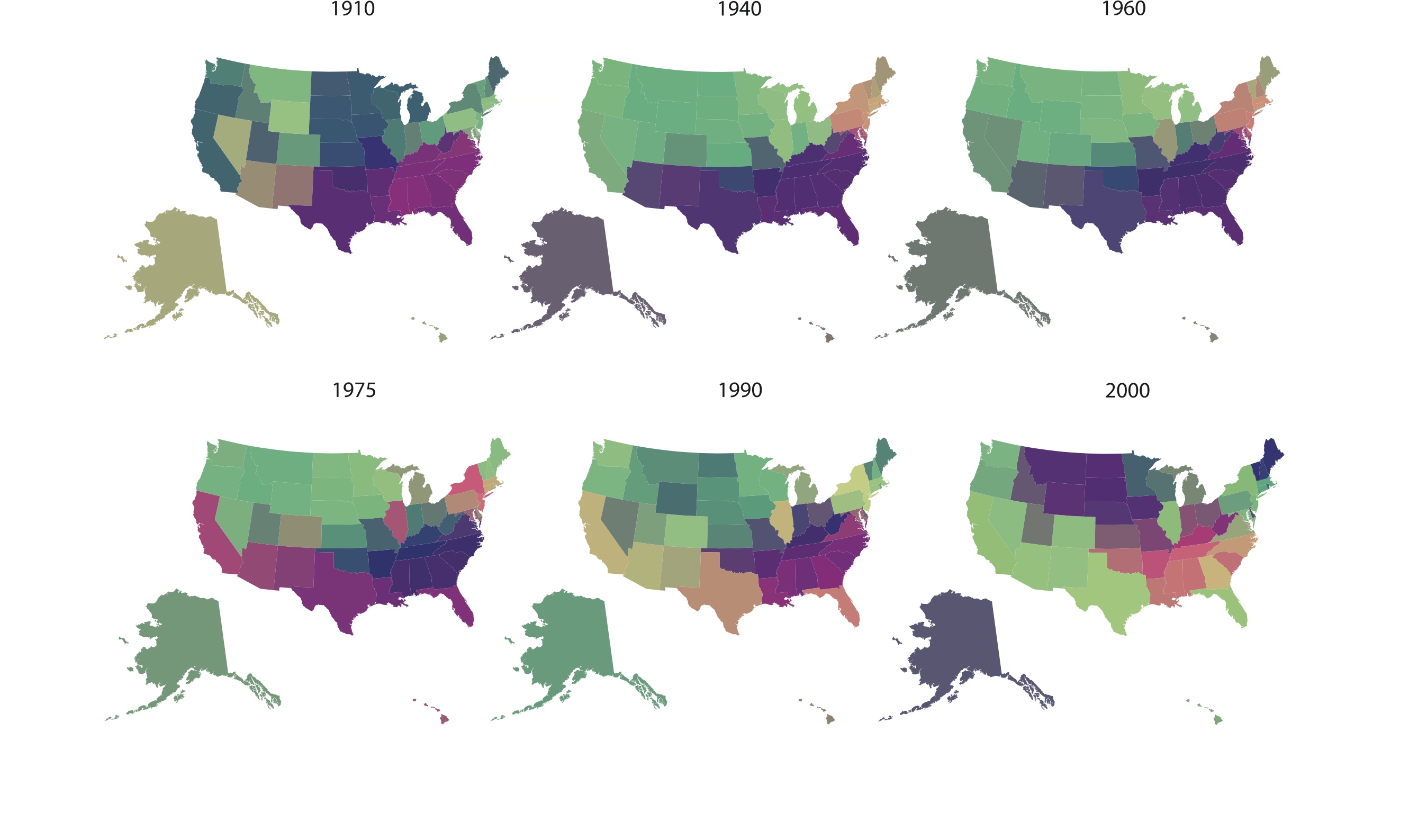}}
\caption{The colors assigned to the states reflect the similarity in their distributions of names. We use a notion of similarity based on the Principal Component Analysis of the matrices of states correlations $C_{ij}$, defined in the text. Details are provided in Methods. A clear difference between the central decades of the twentieth century and its last decades is clearly visible. Northern and southern states were very correlated among them and were forming two separated, uncorrelated entities until 1960. a new configuration emerges at the end of the twentieth century and eventually ends up with a patchy situation where central and coastal states are correlated among them (with very long range correlations covering thousands of miles of physical distance) and are roughly part of two different cultural areas}\label{fig3}
\end{figure*}

We obtain a clear division of states in homogeneous groups. A group of states is qualitatively defined as states that share a certain similarity in their distributions of names. It is natural to associate them to a common cultural area. In Fig. \ref{fig1}, states in the same group are assigned similar colors. This group structure is robust over timescales of the order of a few years: it is thus worth looking at their evolution over larger timescales. While in the beginning of the XX century states were mainly separated between northern and southern states and this separation remains stable across many years, this structure suddenly breaks down in the last decades of the XX century, and a new configuration of groups emerge. The evolution of these groups of states is clear in Fig. \ref{fig3}.
 
\begin{figure*}
\centerline{\includegraphics[width=0.8\textwidth]{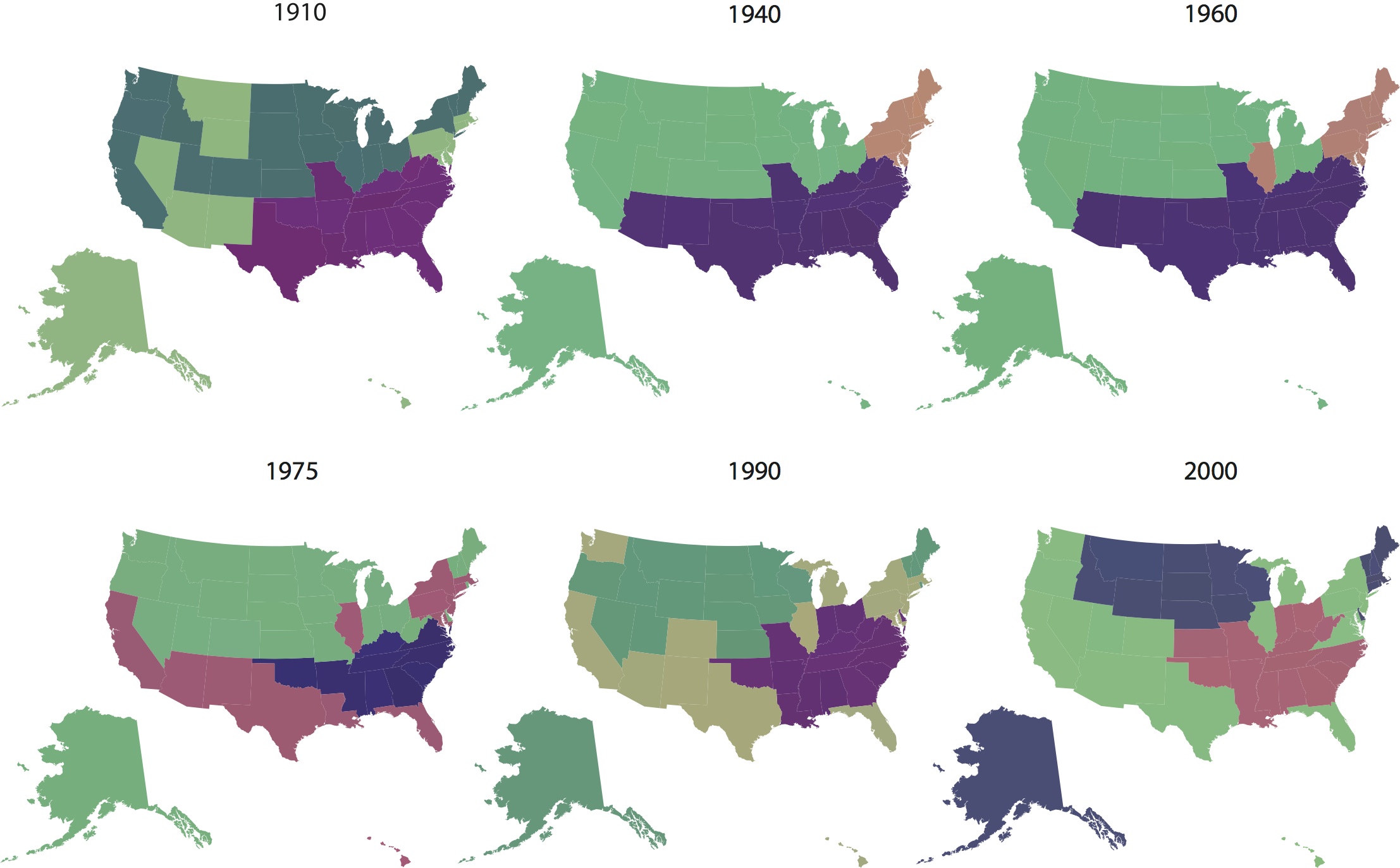}}
\caption{Hierarchical clustering of the states. This analysis clearly shows that groups of states form well separated clusters. Details on the clustering method and explicative movies are provided in the Supplementary Material. The social evolution of geographical correlations are very clear. }\label{fig4}
\end{figure*}

Eventually, in the new configuration that emerges at the end of the XX century, some states of the Atlantic and of the Pacific coasts share common features and belong to the same group, different from that of central states. In order to better identify these groups of states, we used a hierarchical clustering algorithm. A better visualization of this transition can be observed in the Supplementary Videos, which are available online at \url{http://chimera.roma1.infn.it/NAMES/research.html}. This method gives a more quantitative separation with respect to the groups mentioned above, and leads to the formation of clusters, identified by different colors in Fig. \ref{fig4}. The two different methods give the same answer, and make manifest a very interesting social cross-fertilization. This approach is able to describe the emergence of clusters of states through the analysis of the mutual correlations of their newborns names, and extracts interesting information on the evolution of these clusters.

\section{Discussion}

We do not discuss the origin of these correlations, nor the mechanisms according to which names are given to newborn babies. Some attempts in this direction have been done \cite{xi2014cultural}. We also do not study the reasons why there is a reorganization of clusters in the last decades of the XX century, compared to the relatively stable situation of the first half of the century. This is a very interesting issue that surely deserves a more specific study and we leave it open to future investigations. Regarding this point, we observe that irregularities in the retarded cross correlations between the total distribution of the American names were found around the seventies \cite{xi2014cultural}. This effect is probably related to our results. This effect is probably due to the deep cultural transformation occurred in the United Stated after the Vietnam war \cite{xi2014cultural}. 
Our analysis can be adapted to capture which states influence a given state and which ones are influenced by the same state. This amount to identify a directed network which may be relevant for studying time correlations and culture propagation. In the Internet era one can get high statistics data about an extremely large number of behaviours and cultural traits: using our approach on the combined ensemble of such abundant datasets should allow to organise a precise quantitative understanding about the functioning of cultural influences and their evolution.

\section{Supplementary Information}
\subsection{Distribution of names}
\begin{figure*}
\centerline{\includegraphics[width=0.9\textwidth]{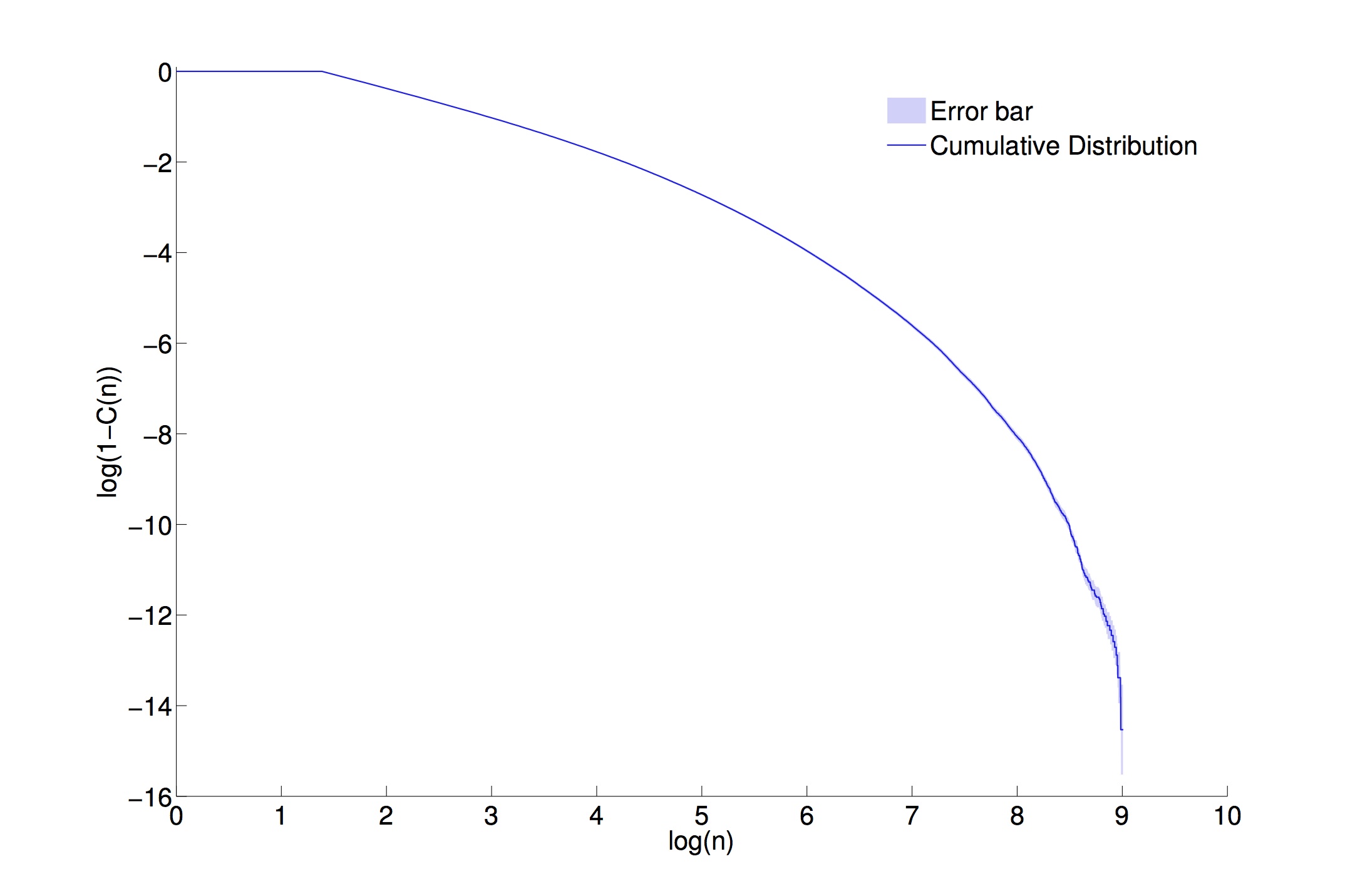}}
\caption{The plateau in the left part of the plot corresponds to the cut-off at $n_{min}=5$, mentioned in the text. The probability distribution of names occurrence has a fat tail and is more complex than a simple power law (that would be a straight line in this plot)
}\label{SIfig1}
\end{figure*}

\begin{figure*}
\centerline{\includegraphics[width=1\textwidth]{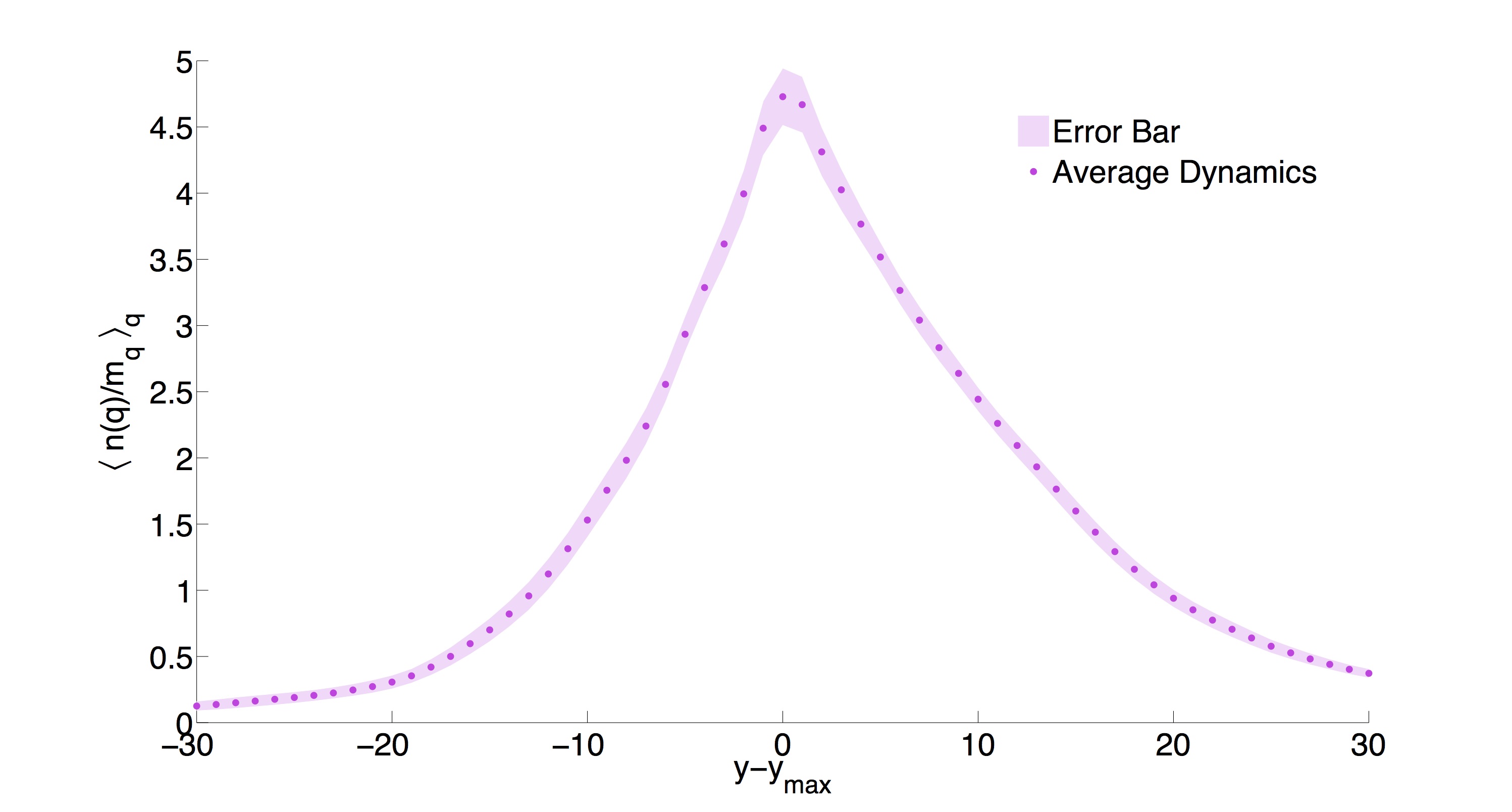}}
\caption{Each name, with rare exceptions, appears, undergoes a rapid ascent and eventually descents to a very low frequency. For each name $q$ we considered the total number of occurrences $n(q)$ in the interval of time ranging from 30 years before to 30 years after the year of the maximum occurrence of q; we averaged over 325 names (we started from the first 1000 most popular names ever given, and we included those which had a peak of popularity between 1940 and 1982, in order to be able to observe both the growth and the decrease of their popularity). We have normalized the number of occurrences by the average occurrences in the whole period ranging from 1910 to 2012, and we have averaged over all these names. On the y-axis $\langle \rangle_q$ stands for the average over different names, while $m_q$ stands for the average value of the occurrences of the name q over the years 1910 to 2012. $y - y_{max}$  in the x-axis stands for the year difference from the peak. 
}\label{SIfig2}
\end{figure*}

The cumulative distribution of the names occurrences, averaged over years, is shown in Fig. \ref{SIfig1}.
While it is evident from the figure that the distribution of names has fat tails, it is also clear that it is
more complex than a power law. It has already been discussed18 that this distribution can be fitted
with a combination of a beta function and a power law.
The vast majority of names appear and spread fast through the states. They stay popular for a few
years, and then disappear (or, more precisely, their frequency goes down to very low, endemic
levels), without keeping a high popularity for a long time. Taking into account many of the most
representative names, their popularity rise and fall can be observed in Fig. \ref{SIfig2}.
Even if this process is not completely symmetric \cite{zanette2012dynamics}, the rate of ascent is very similar to the rate of descent, and the main difference is in the left over tail at very large times (a name that has appeared
does not completely disappear).

\subsection{Comparison with random matrices}
\begin{figure*}
\centerline{\includegraphics[width=.8\textwidth]{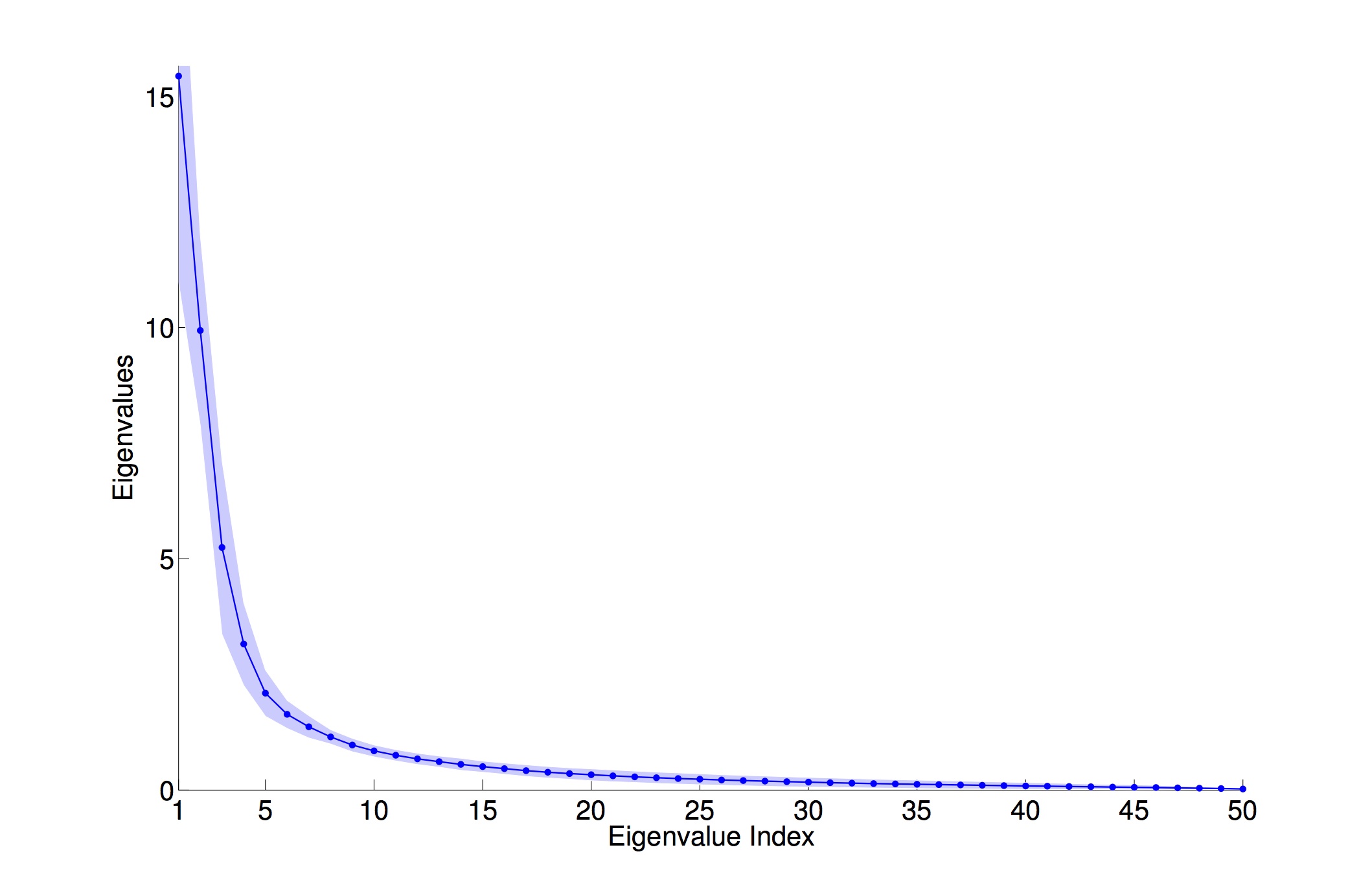}}
\caption{Ordered eigenvalues averaged over the years 1910-2012. Eigenvalues are ordered in a decreasing way, and they are basically stable in time, except at the end of the XX century. The last eigenvalue is always zero and we did not plot it. 
}\label{SIfig3}
\end{figure*}

In all years of the time period under investigation the first eigenvalues of the matrix of correlations
$C_{ij}$ are well separated from all the other eigenvalues, as can be seen in Fig. \ref{SIfig3}.
However, Fig. \ref{SIfig4}. shows that in the last part of the XX century there is no clear
separation between the first and the second eigenvalue. This is not an effect due to the random
noise: this point is rather subtle and we will comment it further in the following section. In this
section we discuss how to deal with the noise, and how to identify system-specific, non random,
correlations. One of the possible methods is to compare the spectrum of the correlation matrices
with that of random Wishart-Laguerre matrices \cite{plerou1999universal}. This method provides bounds for the random bulk of the spectrum and thus the eigenvalues (and the corresponding eigenvectors) outside this interval are thought to yield information on the genuine correlations of the underlying system.
The largest eigenvalues have been shown to play an important role in many situations \cite{plerou2002random}, \cite{edelman1988eigenvalues} and we will see below
that this is the case also in our problem. We used a slightly different method to establish that noise is
not affecting our findings, comparing the spectrum of correlation matrices with the one of random
correlation matrices obtained as follows: in each year, we made a random permutation of the
occurrences of the $N_f$ names inside each state, and we used it to compute the correlation matrix $C_R$, repeating all the steps described in the equations going from eq. (\ref{eq:uno}) to eq. (\ref{eq:cij}). Given that $N f \sim 10^4$ , the correlation matrices $C_R$ are similar to identity matrices, whose spectrum is
obviously made by ones. The spectrum of $C_R$ is different from that of $C$, even if the widely
used unfolded nearest neighbour spacing distribution (see for example \cite{plerou1999universal}) is similar. In order to rule out a role of the random noise, we compared the Inverse Participation Ratio (IPR) $I(\lambda)$ of our data to that of the null model. This test provides a direct evaluation of the non-random part of the spectrum. It is defined as follows: be $v^{\lambda}_i$ the $i$-th component of the (normalized) eigenvector $v^{\lambda}$ such that $C v^{\lambda} = \lambda v^{\lambda} $; thus 
\begin{equation}
I(\lambda)=\sum_{i=1}^M \left[v^{\lambda}_i\right]^4\:
\end{equation}
where $M$ is the total number of components. This quantities is often used in localization problems, since its reciprocal $1/I(\lambda)$ equals the number of states over which a vector is localized. An easy way to understand this point is to use the normalization condition $\sum_{i} v_i^2=1$, from which we see that $v_i \sim 1/\sqrt{c} $ for a vector localized over $c$ states. Thus, we see that $I(\lambda) \sim 1/c$. A localized vector is a vector for which $c \sim 1$, while for non-localized vectors $c \sim O(M)$. 
We compared the IPR of all the eigenvectors of $C_{ij}$ and $C^R_{ij}$ and we found that while there is a complete agreement for the respective $I(\lambda)$ in the right region of the spectrum (small eigenvalues), there is a clear separation in the left part of the spectrum, which can be interpreted as a deviation from the random matrices behaviour, as can be seen in Fig. \ref{SIfig5}.
The first eigenvalues carry relevant information for the detection of collective modes in the system:
they are found to be non-localized, and to give important information on the correlation between
states. The fact that the distribution of names occurrences is fat-tailed is crucial for the qualitative
agreement we find in the right region of the spectrum, as it is known that, in cross-correlation
matrices of signals sampled from a fat-tailed distribution, eigenvectors corresponding to small
eigenvalues are found to be localized \cite{cizeau1994theory}, while this does not happen in the Gaussian case, where the IPR is flat $I(\lambda)\sim 3/M$ for all $\lambda$ \cite{plerou1999universal}, \cite{barthelemy2002large}, \cite{edelman1988eigenvalues}.

\subsection{Principal component analysis and hierarchical clustering}

Let us introduce Principal Component Analysis (PCA) and hierarchical clustering, which are the two methods we used to investigate the genuine correlations of the states through the years.
A simple way to extract informations from the correlation matrix $C$ is the Principal Component Analysis (PCA) \cite{Johnstone}, which consists in a partial eigen-decomposition of the correlation matrix defined in eq. (\ref{eq:cij}).  It uses just the largest eigenvalues and their corresponding eigenvectors to reconstruct collective modes of the system.
It can be described by means of an example. Suppose that the correlation matrix is such that the non-diagonal elements of the line $i$ are all very small except for a single element $C_{ij}$, which is almost one, and that all the other non diagonal terms are also very small (but $C_{ji}=C_{ij}$).
Quite intuitively, this situation corresponds to some kind of connection between $i$ and $j$.
In this case the eigenvector of $C$ that corresponds to the largest eigenvalue contains this information in its components: indeed, it can be shown that the only component of $v$ significantly different from zero are $v_i$ and $v_j$.
Now suppose that the correlation matrix is a bit more complicated, and that many non-diagonal elements are large. Although, in some situations this effect is purely due to the noise, often it may happen because complicated pattern are hidden in the system, more complex than pairwise correlations, which involve many states.
It is quite reasonable that eigenvectors of $C$ take into account this information, and eigenvectors corresponding to largest eigenvalues are localized on such patterns.
Thus PCA may led to a reconstruction of the complex correlations among many states.
In order to get rid of the noise contained in the correlation matrices, and to be sure that eigenvectors corresponding to large eigenvalues contains genuinely non-random informations, we proceeded as explained in the previous section.

Taking the first two eigenvectors of $C$, $v^{(1)}$ and $v^{(2)}$, a bi-dimensional representation of these eigenvectors can be obtained by plotting the $M=51$ points of coordinates $(v_i^{(1)},v_i^{(2)})$, one for each state. This can be done year after year and the result is shown in Fig. \ref{SIfig8}.
In these scattered plots, clusters/groups of states that are similar, in a sense that will be discussed below, naturally emerge.

The notion of similarity can be understood with the following qualitative description. 
Let us define $\widehat{C}=(1+C)/2$, which is a positive definite matrix, whose elements satisfy $0 <\widehat{C}_{ij}<1$. Let us assume that the state $i$ has large correlation with a group of states $\mathcal{N}(i)$, while it has small correlations with the rest of the states, i.e.\ $\widehat{C}_{ij}\simeq 0$ when $j \notin \mathcal{N}(i)$.
Let us also assume that another state $k$ has a large correlation with a group of states $\mathcal{N}(k)$ whose overlap with $\mathcal{N}(i)$ is big enough.
In this case, $i$ and $k$ are said to be similar, even if the correlation coefficient $\widehat{C}_{ik}$ between them is small, and those two states belong to the same cluster. 

These results have been compared with those obtained using an agglomerative hierarchical clusterization based on the matrix $\widehat{C}$. Hierarchical clustering is a method to build a hierarchy of clusters starting from $M$ clusters of one state each and such that clusters are merged as one moves up in the hierarchy.
This can be done in many ways.
First one has to define a metric $d(i,j)$, for example the Euclidean one, between the columns of the correlation matrix.
Then we have to minimize a certain distance $D(X,Y)$ between the clusters $X$ and $Y$:  at each step, the two clusters separated by the shortest distance are merged.
There are several possible definitions of $D(X,Y)$, each one defining a possible criterion to perform the clusterization. 
We used the \textit{complete} criterion, which is defined by 
\begin{equation}
D(X,Y)=\max_{i\in X, y \in Y} d(i,j)
\end{equation}
which tends to find compact clusters of approximately equal diameters. This procedure leads to the construction of a tree diagram called \textit{dendrogram}, which can be cut at the height one prefers. This introduces a sort of arbitrariness in the number of clusters that are considered. Looking at the distribution of the scatter plot in Fig. \ref{SIfig6}, we decided to fix the number of clusters to 3.

It can be noticed that states that are similar in the sense specified before, are found to belong the same cluster.
In each year, in order to measure of the quality of the clusterization obtained, we computed the cophenetic coefficient. 
Given the Euclidean distance $E(i,j)$ between the points $i$ and $j$, and the corresponding distance that these points have on the dendrogram $D(i,j)$, cophenetic coefficient $c$ can be defined as 
\begin{equation}
c = \frac {\sum_{i<j} (E(i,j) - \bar{E})(D(i,j) - \bar{D})}{\sqrt{\left[\sum_{i<j}(E(i,j)-\bar{E})^2\right] \left[\sum_{i<j}(D(i,j)-\bar{D})^2\right]}}.
\label{copheneticcoeff}
\end{equation}
where $\bar{E}$ and $\bar{D}$ are, respectively, the average values of the Euclidean and of the dendrogramatic distances. 
Values of $c$ close to one, indicate a good clusterization, since the correlation coefficient between the actual distances and the dendrogrammatic one is large. 
In each year, $c$ is found to be near $0.85$ and never smaller that $0.7$. PCA and hierarchical clustering results can be be found in Fig. \ref{fig3} and in Fig. \ref{fig4} of the main text.

\subsection{What are the principal components made of}
\begin{figure*}
\centerline{\includegraphics[width=.9\textwidth]{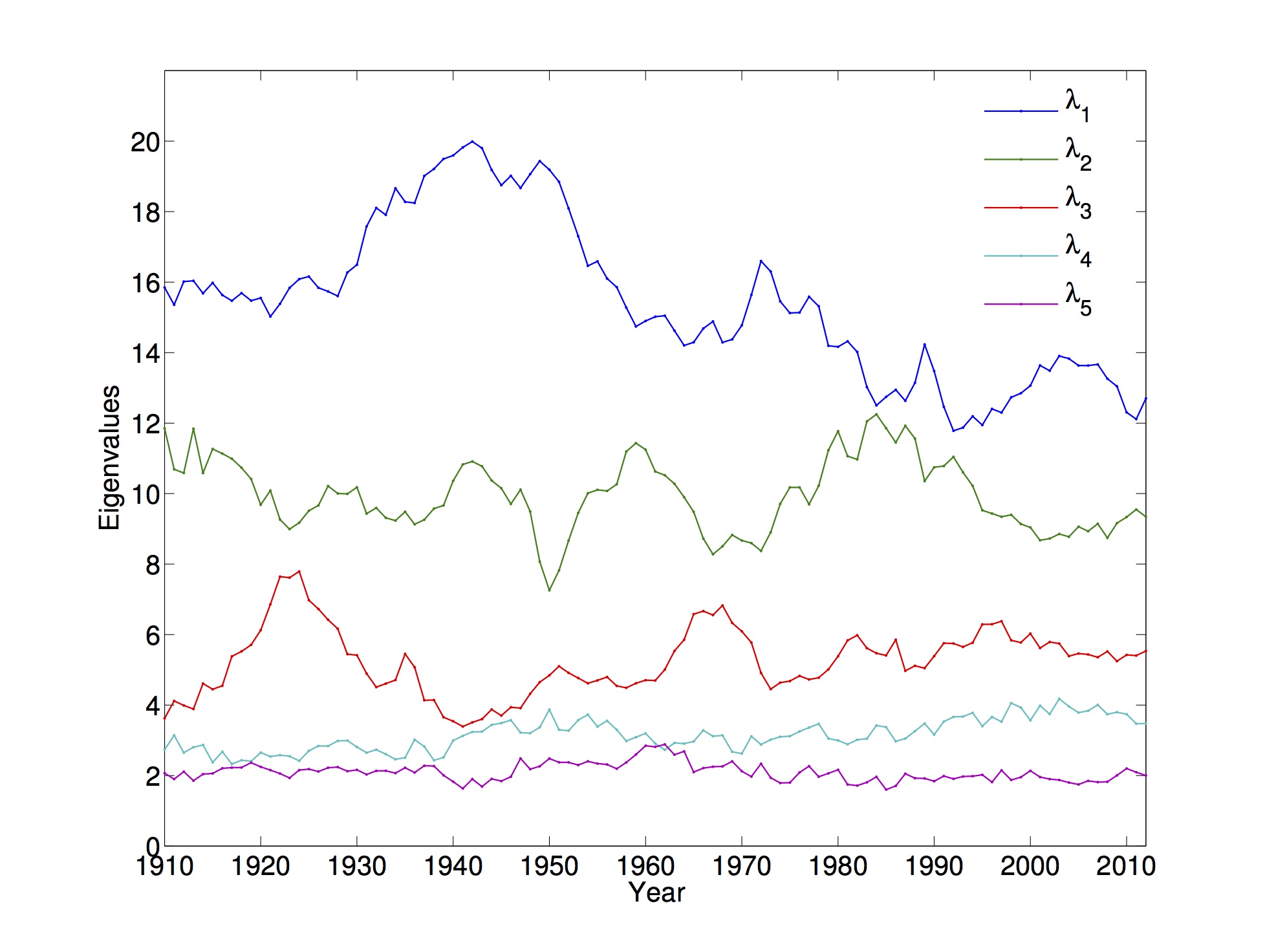}}
\caption{The first 5 eigenvalues of the correlation matrix plotted versus time. The first two eigenvalues are close to each other in the last part of the XX century. 
}\label{SIfig4}
\end{figure*}

\begin{figure*}
\centerline{\includegraphics[width=.9\textwidth]{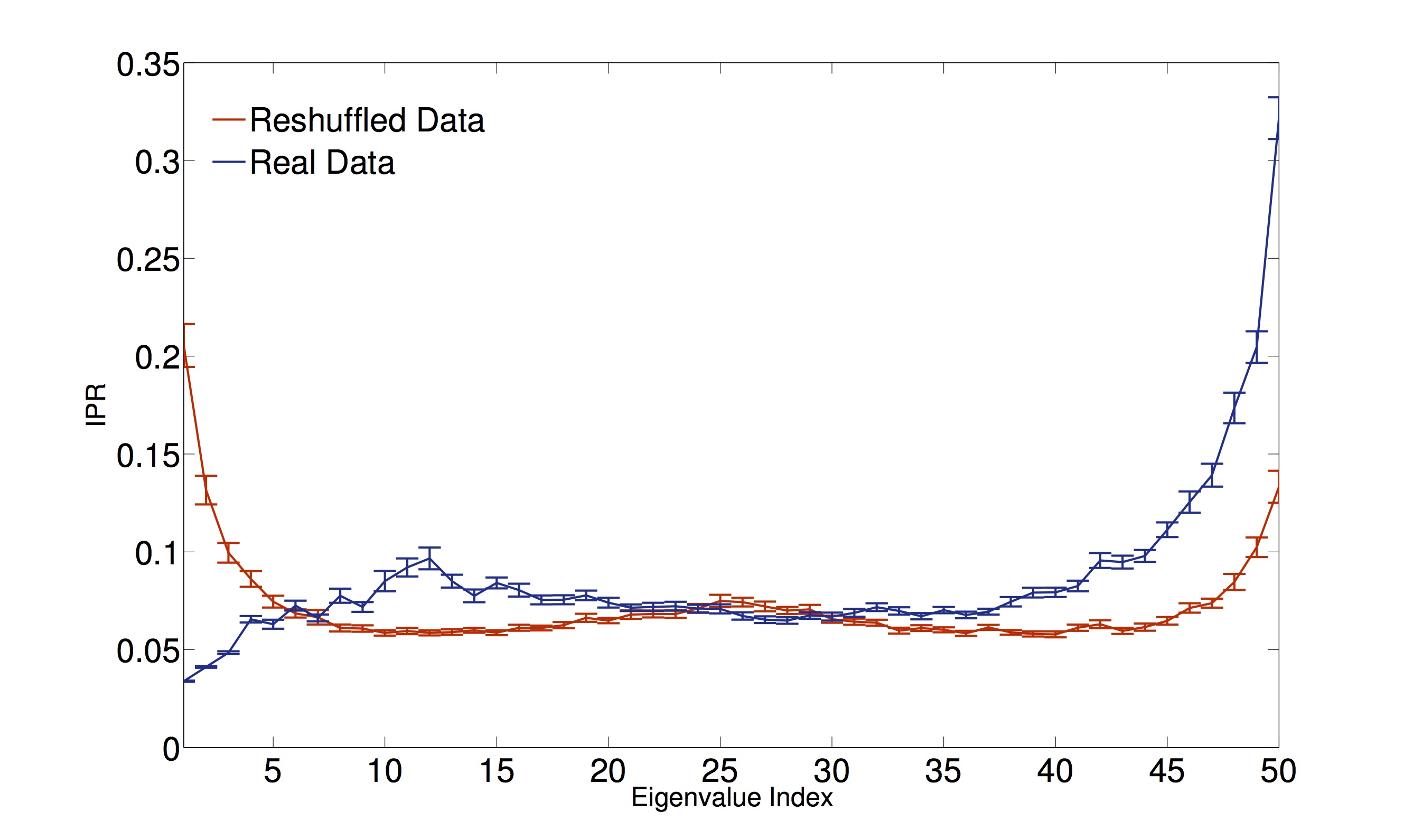}}
\caption{Mean IPR computed for every eigenvalue from $C$ (blue line) and from $C_R$ (red line) averaging the respective IPR in different years. $C_R$ plays the role of a very simple null-model. Eigenvalues are ordered in a decreasing way. An unambiguous deviation from the random matrices IPR behaviour is found in the left part of the spectrum, i.e. for the largest eigenvalues. 
}\label{SIfig5}
\end{figure*}

\begin{figure*}
\centerline{\includegraphics[width=.9\textwidth]{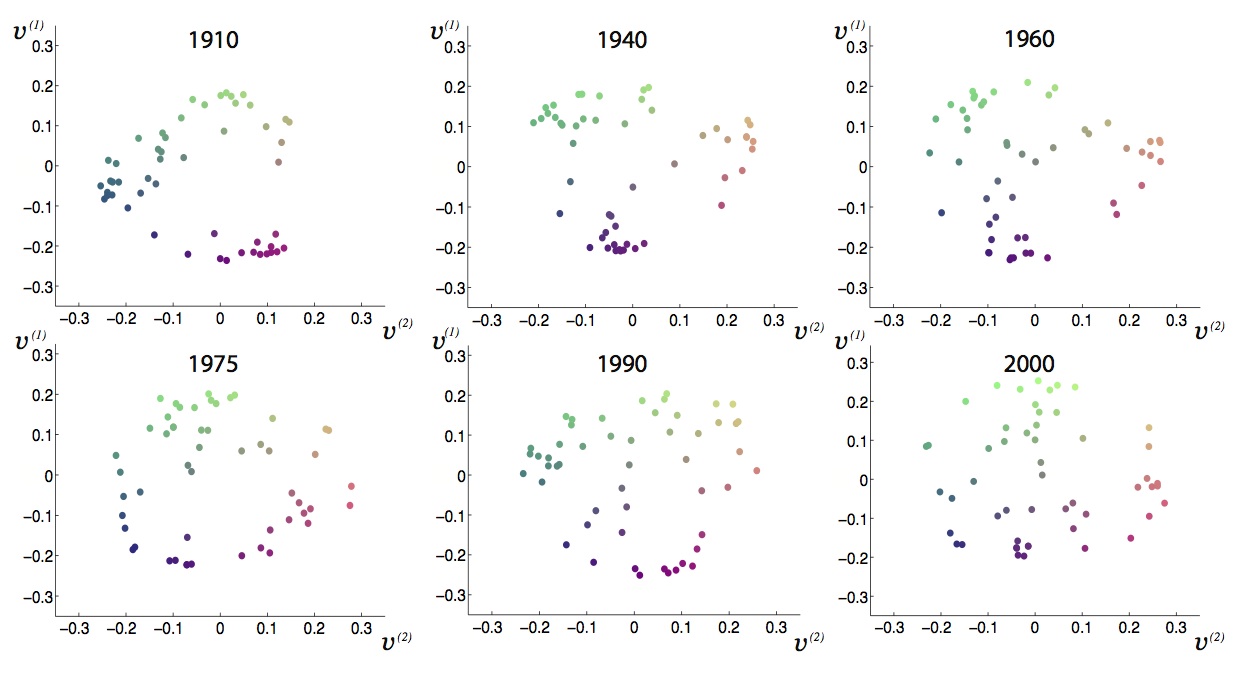}}
\caption{Projections of the 51 states onto the plane made by the first two eigenvectors of C, $v^2$ and $v^2$. Colors are assigned according to the position in this plane, and these colors are exactly the colors used in Fig. 3 of the main text. 
}\label{SIfig6}
\end{figure*}

\begin{figure*}
\centerline{\includegraphics[width=.8\textwidth]{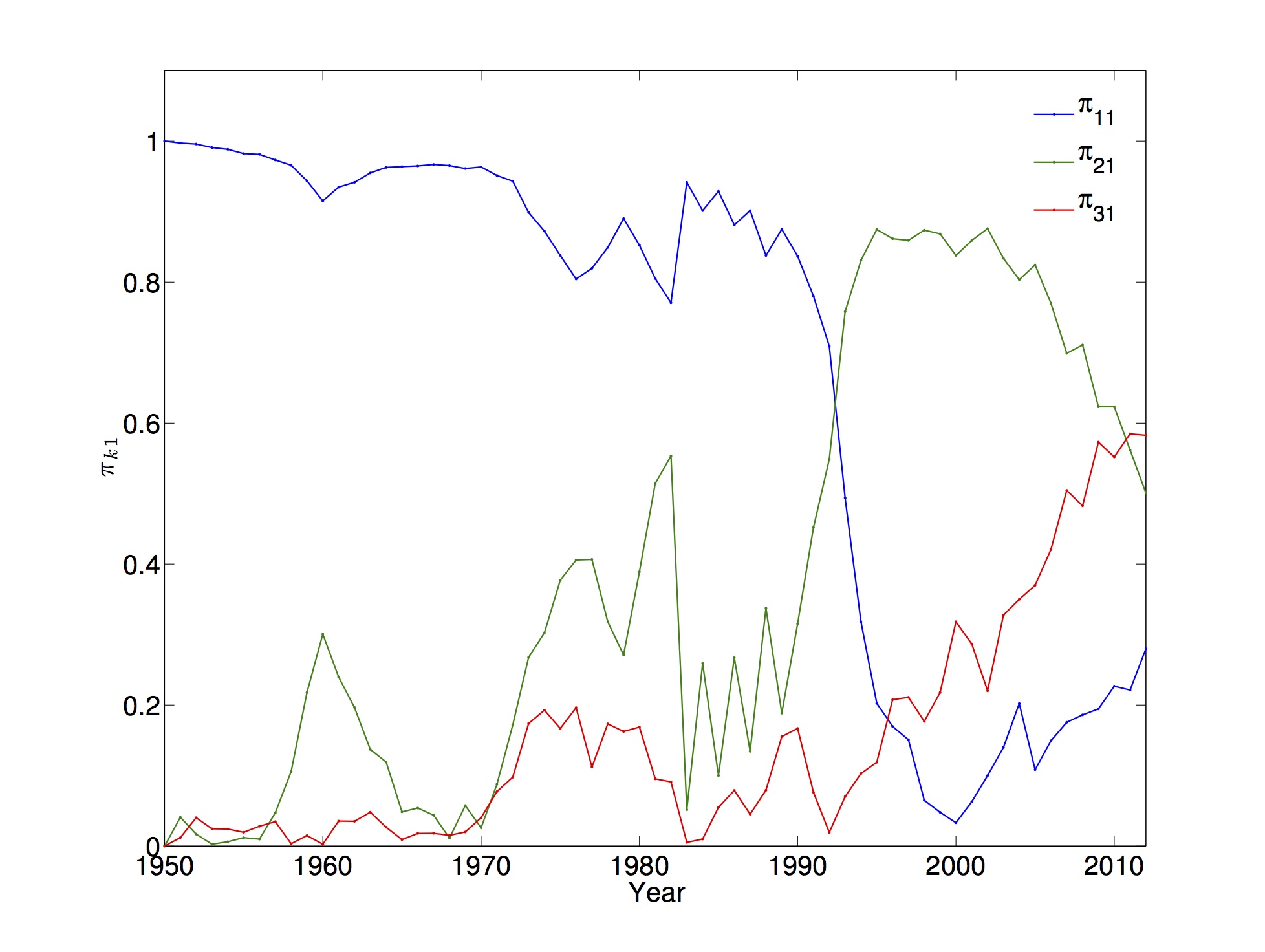}}
\caption{Projections $\pi_{k1}$ of the first three eigenvectors $(k = 1, 2, 3)$ of C(Year) on the first eigenvector of $C(1950)$. In 1992-1993 there is a crossing point between $\pi_{21}$ and $\pi_{11}$. Thus the first principal component during the years 1950-1990 becomes the second principal component in the following years. 
}\label{SIfig7}
\end{figure*}

\begin{figure*}
\centerline{\includegraphics[width=.8\textwidth]{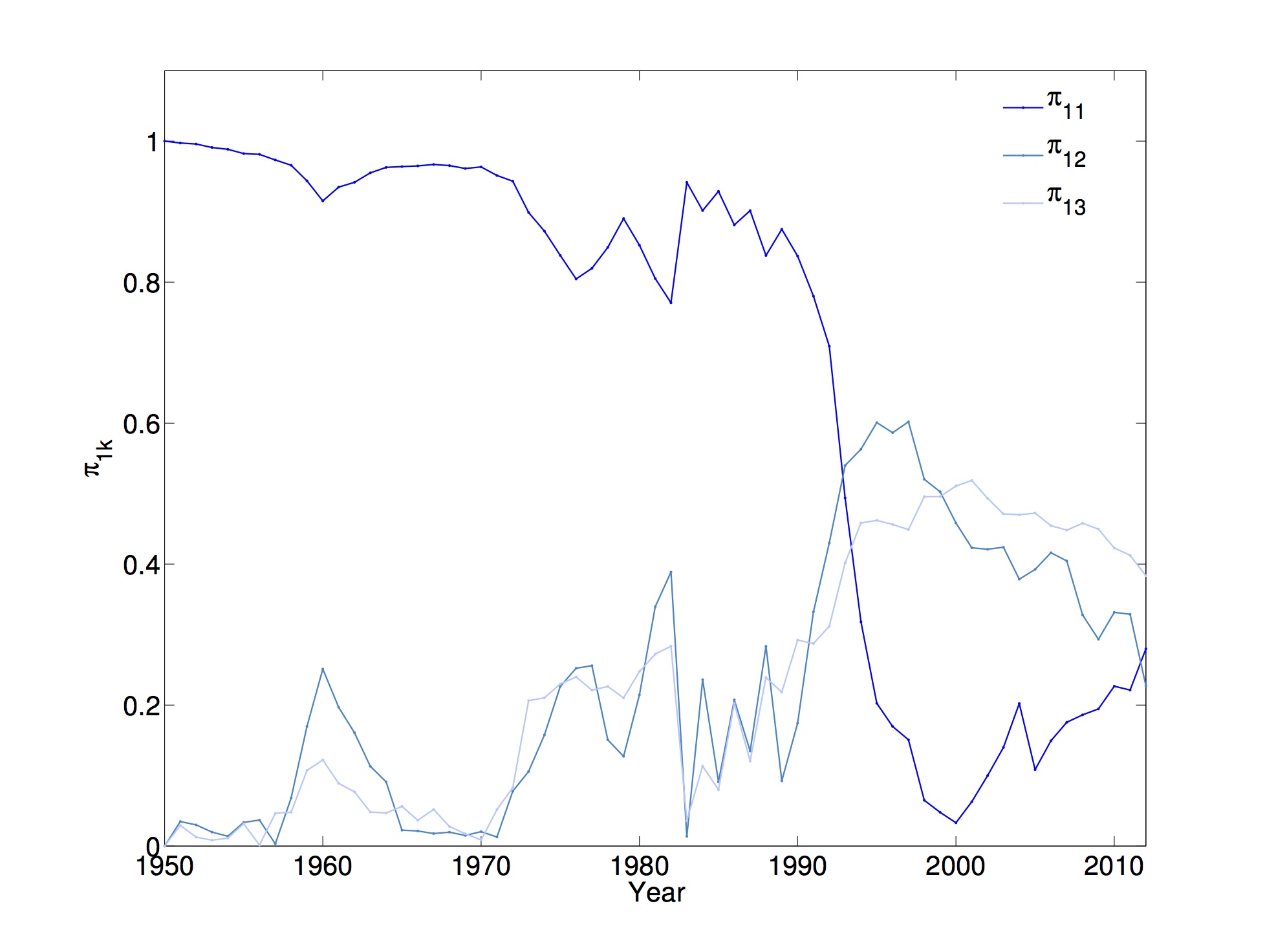}}
\caption{Projections $\pi_{1k}$ of the first eigenvector of C(Year) on the first three eigenvectors $(k = 1, 2, 3)$ of $C(1950)$. After 1993 the principal component is mainly a superposition of the second and third eigenvectors of the 1950 correlation matrix, even though recently $\pi_{11}$ has grown up again.
}\label{SIfig8}
\end{figure*}

As we noticed in Fig. \ref{SIfig4} the first eigenvalues are always well separated during the first decades of the XX century, indicating persisting patterns in the states that survive during years.
We also noticed that the situation is less clear at the end of the XX century, and there is not a clear separation between the first two eigenvalues. Here we will further discuss this point.
In each year, we considered the first two eigenvectors of the correlation matrix $C_{ij}$ and we assigned a color to each state, corresponding to the position it had in the plane spanned by $v^{(1)}$ and $v^{(2)}$, i.e. the eigenvectors corresponding to the two largest eigenvalues.
While these persistent patterns are quite clear in the first decades, indicating a separation between northern states and southern states, we also notice that the situation changes at the end of the XX century, when states start to be grouped between coastal states, and central states (a better visualization of such a transition can be obtained from the movies available on the website \url{http://chimera.roma1.infn.it/NAMES/research.html} that show data for all the years).
In order to better understand what has happened in the transition years, we studied how eigenvectors changed during years. 
Fig. \ref{SIfig4} suggests to choose, as a reference, the eigenvectors of the correlation matrix in 1950, since it is the year with the largest difference $\lambda_1 - \lambda_2$.
We then studied how the first three eigenvectors of the correlation matrix in the year $y$ are related to those of 1950.
For each year we evaluated the projections of $k$-th eigenvector on $j$-th 1950's eigenvector 
\begin{equation}
\pi_{kj}(y)=\sum_{i=1}^M u_i^k(y)u_i^j(1950).
\end{equation}
Observing these projections we see that the first eigenvector is rather stable until nineties, when it becomes mostly a superposition of 1950's second and third eigenvectors. Conversely, after this transition, the second eigenvector is mainly constituted by the first 1950's eigenvector, as can be seen in Fig. \ref{SIfig7} and Fig. \ref{SIfig8}.
These figures provide a clear interpretation of the evolution of the correlations between states: for many decades of the XX century there has been a stable configuration of groups (clusters) of states that eventually broke down at the end of the century. We do not further investigate the reasons behind this change. Looking at the distribution of baby names is already enough to make a sensible and realistic clusterization of states emerge, and this clusterization is related to cultural influences. The deep motivations that lie behind this change deserve to be studied by the sociologists, maybe helped by further quantitative testing of large data sets. 

\clearpage

\begin{thebibliography}{10}

\bibitem{belleau1987cyclical}
Bonnie~D Belleau.
\newblock Cyclical fashion movement: women's day dresses: 1860-1980.
\newblock {\em Clothing and textiles research journal}, 5(2):15--20, 1987.

\bibitem{bentley2007regular}
R~Alexander Bentley, Carl~P Lipo, Harold~A Herzog, and Matthew~W Hahn.
\newblock Regular rates of popular culture change reflect random copying.
\newblock {\em Evolution and Human Behavior}, 28(3):151--158, 2007.

\bibitem{herzog2004random}
Harold~A Herzog, R~Alexander Bentley, and Matthew~W Hahn.
\newblock Random drift and large shifts in popularity of dog breeds.
\newblock {\em Proceedings of the Royal Society of London. Series B: Biological
  Sciences}, 271(Suppl 5):S353--S356, 2004.

\bibitem{neiman1995stylistic}
Fraser~D Neiman.
\newblock Stylistic variation in evolutionary perspective: inferences from
  decorative diversity and interassemblage distance in illinois woodland
  ceramic assemblages.
\newblock {\em American Antiquity}, pages 7--36, 1995.

\bibitem{bentley2008random}
R~Alexander Bentley.
\newblock Random drift versus selection in academic vocabulary: An evolutionary
  analysis of published keywords.
\newblock {\em PloS one}, 3(8):e3057, 2008.

\bibitem{krawczyk2014simmel}
Malgorzata~J Krawczyk, Antoni Dydejczyk, and Krzysztof Kulakowski.
\newblock The simmel effect and babies' names.
\newblock {\em Physica A: Statistical Mechanics and its Applications},
  395:384--391, 2014.

\bibitem{xi2014cultural}
Ning Xi, Zi-Ke Zhang, Yi-Cheng Zhang, Zehui Ge, Li~She, and Kui Zhang.
\newblock Cultural evolution: The case of babies' first names.
\newblock {\em Physica A: Statistical Mechanics and its Applications},
  406:139--144, 2014.

\bibitem{berger2012karen}
Jonah Berger, Eric~T Bradlow, Alex Braunstein, and Yao Zhang.
\newblock From karen to katie using baby names to understand cultural
  evolution.
\newblock {\em Psychological science}, 23(10):1067--1073, 2012.

\bibitem{jensen1998self}
Henrik~Jeldtoft Jensen.
\newblock {\em Self-organized criticality: emergent complex behavior in
  physical and biological systems}, volume~10.
\newblock Cambridge university press, 1998.

\bibitem{albert2002statistical}
R{\'e}ka Albert and Albert-L{\'a}szl{\'o} Barab{\'a}si.
\newblock Statistical mechanics of complex networks.
\newblock {\em Reviews of modern physics}, 74(1):47, 2002.

\bibitem{feldman1981further}
Marcus~W Feldman and Luca~L Cavalli-Sforza.
\newblock Further remarks on darwinian selection and "altruism".
\newblock {\em Theoretical Population Biology}, 19(2):251--260, 1981.

\bibitem{zanette2012dynamics}
Damian~H Zanette.
\newblock Dynamics of fashion: The case of given names.
\newblock {\em arXiv preprint arXiv:1208.0576}, 2012.

\bibitem{acerbi2012logic}
Alberto Acerbi, Stefano Ghirlanda, and Magnus Enquist.
\newblock The logic of fashion cycles.
\newblock {\em PloS one}, 7(3):e32541, 2012.

\bibitem{berger2009adoption}
Jonah Berger and Gael Le~Mens.
\newblock How adoption speed affects the abandonment of cultural tastes.
\newblock {\em Proceedings of the National Academy of Sciences},
  106(20):8146--8150, 2009.

\bibitem{Simmel}
Simmel G.
\newblock Fashion.
\newblock {\em The American Journal of Sociology}, 62(6):541--558, 1904.

\bibitem{hahn2003drift}
Matthew~W Hahn and R~Alexander Bentley.
\newblock Drift as a mechanism for cultural change: An example from baby names.
\newblock {\em Proceedings of the Royal Society of London. Series B: Biological
  Sciences}, 270(Suppl 1):S120--S123, 2003.

\bibitem{ssa}
{\em The Official Website of the US Social Security Administration
  www.ssa.gov}.

\bibitem{li2012analyses}
Wentian Li.
\newblock Analyses of baby name popularity distribution in us for the last 131
  years.
\newblock {\em Complexity}, 18(1):44--50, 2012.

\bibitem{sporns2004organization}
Olaf Sporns, Dante~R Chialvo, Marcus Kaiser, and Claus~C Hilgetag.
\newblock Organization, development and function of complex brain networks.
\newblock {\em Trends in cognitive sciences}, 8(9):418--425, 2004.

\bibitem{luo2006application}
Feng Luo, Jianxin Zhong, Yunfeng Yang, Richard~H Scheuermann, and Jizhong Zhou.
\newblock Application of random matrix theory to biological networks.
\newblock {\em Physics Letters A}, 357(6):420--423, 2006.

\bibitem{luo2007constructing}
Feng Luo, Yunfeng Yang, Jianxin Zhong, Haichun Gao, Latifur Khan, Dorothea~K
  Thompson, and Jizhong Zhou.
\newblock Constructing gene co-expression networks and predicting functions of
  unknown genes by random matrix theory.
\newblock {\em BMC bioinformatics}, 8(1):299, 2007.

\bibitem{jalan2010random}
Sarika Jalan, Norbert Solymosi, G{\'a}bor Vattay, and Baowen Li.
\newblock Random matrix analysis of localization properties of gene
  coexpression network.
\newblock {\em Physical Review E}, 81(4):046118, 2010.

\bibitem{laloux1999noise}
Laurent Laloux, Pierre Cizeau, Jean-Philippe Bouchaud, and Marc Potters.
\newblock Noise dressing of financial correlation matrices.
\newblock {\em Physical Review Letters}, 83(7):1467, 1999.

\bibitem{plerou1999universal}
Vasiliki Plerou, Parameswaran Gopikrishnan, Bernd Rosenow, Lu{\'\i}s A~Nunes
  Amaral, and H~Eugene Stanley.
\newblock Universal and nonuniversal properties of cross correlations in
  financial time series.
\newblock {\em Physical Review Letters}, 83(7):1471, 1999.

\bibitem{plerou2002random}
Vasiliki Plerou, Parameswaran Gopikrishnan, Bernd Rosenow, Luis A~Nunes Amaral,
  Thomas Guhr, and H~Eugene Stanley.
\newblock Random matrix approach to cross correlations in financial data.
\newblock {\em Physical Review E}, 65(6):066126, 2002.

\bibitem{conlon2009cross}
Thomas Conlon, Heather~J Ruskin, and Martin Crane.
\newblock Cross-correlation dynamics in financial time series.
\newblock {\em Physica A: Statistical Mechanics and its Applications},
  388(5):705--714, 2009.

\bibitem{barthelemy2002large}
Marc Barth{\'e}lemy, Bernard Gondran, and Eric Guichard.
\newblock Large scale cross-correlations in internet traffic.
\newblock {\em Physical Review E}, 66(5):056110, 2002.

\bibitem{santhanam2001statistics}
MS~Santhanam and Prabir~K Patra.
\newblock Statistics of atmospheric correlations.
\newblock {\em Physical Review E}, 64(1):016102, 2001.

\bibitem{edelman1988eigenvalues}
Alan Edelman.
\newblock Eigenvalues and condition numbers of random matrices.
\newblock {\em SIAM Journal on Matrix Analysis and Applications},
  9(4):543--560, 1988.

\bibitem{cizeau1994theory}
P~Cizeau and JP~Bouchaud.
\newblock Theory of l{\'e}vy matrices.
\newblock {\em Physical Review E}, 50(3):1810, 1994.

\bibitem{Johnstone}
 Iain M. Johnstone.
 \newblock  High dimensional statistical inference and random matrices.
 \newblock {\em EuropeanMathematical Society}, (2007): 307.


\end{thebibliography}
\end{document}